\documentclass[dvips]{article}

\usepackage{icrc2011}

\title{Field test of the hybrid photodetector R9792U-40 on the MAGIC camera}

\newcommand{\etal}{\MakeLowercase{\textit{et al. }}} 
\shorttitle{T.Y. Saito \etal Field test of HPD R9792U-40 on the MAGIC camera}

\authors{T.Y. Saito$^{1}$, S. Sun$^{1}$, R. Orito$^{1}$, E. Lorenz$^{2}$, R. Mirzoyan$^{1}$, M. Teshima$^{1,3}$, M. Garczarczyk$^{4}$\\
on behalf of the MAGIC Collaboration}
\afiliations{
$^1$Max-Planck-Institut f\"ur Physik, M\"unchen, Germany \\
$^2$ETH z\"urich, Switzerland\\ 
$^3$Institute for Cosmic Ray Research, Kashiwa, Japan\\
$^4$Instituto de Astrofisica de Canarias, La Laguna, Spain 
 }
\email{tysaito@mppmu.mpg.de}

 \abstract{
The hybrid photodetector (HPD) R9792U-40 has very high peak quantum efficiency ($>50$\% at 500 nm), excellent charge resolution and very low after-pulsing probability (500 times less than that of currently used photomultipliers (PMTs)).
These features will improve the sensitivity, the energy resolution and the energy threshold of the MAGIC telescope.
On the other hand, its high photocathode voltage ($-8$ to $-6$ kV), relatively short photocathode lifetime, and relatively large temperature dependence of the gain need to be taken care of. 
In February 2010, 6 HPDs were installed in a corner of the MAGIC-II camera for a field test.
Here we report the results of the field test and our future plans.
}

\begin{document}
\maketitle

\section{Introduction}

The MAGIC telescope is a new generation Imaging Atmospheric Chereknkov Telescope (IACT) system, located on the Canary Island of La Palma (27.8$^\circ$ N, 17.8$^\circ$W, 2225 m asl). It consists of two telescopes with 
the reflector diameter of 17~m. The reflector size is the largest among the currently operating IACTs. It is one of the key parameters, which result that MAGIC has the lowest energy threshold of 60 GeV.
The low energy threshold is important for the following reasons:
(a) The extragalactic background light absorbs high energy gamma-rays. 
As the energy of the gamma-ray goes lower, the Universe becomes less opaque, 
which allowes to see farther gamma-ray sources like distant AGNs or GRBs. (b) 
Source types like pulsars have very steep spectrum above 10 GeV. A low energy threshold of the detector is crucial for their studies.

In order to lower further the energy threshold and improve the sensitivity below 100 GeV in MAGIC,
detection of  more Cherenkov photons from air showers is necessary. As it is impossible to enlarge the reflector diameter of a telescope retrospectively, another solutions are under investigation. One of the solutions is the improvement of the photo-detection efficiency of the photosensors in the camera.
For this reason we developed together with HAMAMATSU Photonics a new HPD R9792U-40.
The field test of this photosensor started in February 2010 and its results are reported here.

\section{HPD R9792U-40}

\subsection{Structure and working principle}
HPD R9792U-40 consists of a GaAsP photocathode, focusing electrodes and an avalanche photodiode (APD). The photocathode has a round shape with a diameter of 18~mm. The APD has a cylindrical shape with a diameter of 3~mm. Two voltages are applied to the HPD: $-8 \sim -6$~kV to the photocathode and $300 \sim 450$~V to the APD anode (inverse bias voltage).
When photons hit the photocathode, photoelectrons (ph.e's) are produced. They are accelerated in the strong electromagnetic field and bombard the APD.
This bombardment produces $\sim 1500$ electron-hole pairs per one ph.e. (in the case of $\sim -8$kV photocathode voltage) in the depleted layer of the APD.
The secondary electrons are subsequently accelerated in the high electric field of the pn structure of the APD and
initiate avalanches, providing an additional gain of $30 \sim 50$. 

\subsection{Advantages and disadvantages}
\begin{figure}[h]
\centering
  \includegraphics[width=2.in]{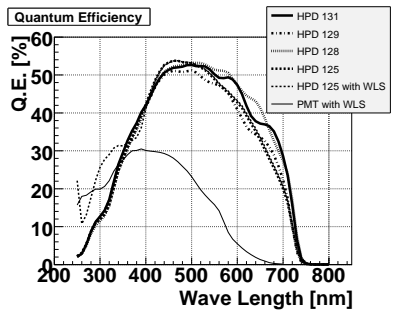}
  \caption{QE curve of several HPD's R9792U-40. In addition the QE curve of one of the currently used PMT's
is also shown as a reference.}
  \label{FigQE}
 \end{figure}

\begin{figure}[h]
\centering
   \includegraphics[width=2.in]{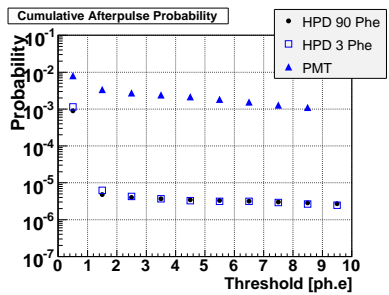} 
  \caption{Afterpulsing rate of the HPD R9792U-40 as a function of the threshold.
Above 1.5 ph.e. level the rate is 500 times lower than the currently used PMTs.}
   \label{FigAP}
 \end{figure}

Compared to the conventional PMTs, following advantages and disadvantages can be listed up for HPDs: \\
{\bf Advantages}
\begin{description}
\item [1)] Twice higher photo-detection efficiency (see Fig. \ref{FigQE}).
\item [2)] 500 times lower afterpulsing rate (see Fig. \ref{FigAP}).
\item [3)] Excellent charge resolution (see \cite{RefHPD}).
\end{description}

{\bf Disadvantages}
\begin{itemize}
\item [4)] Higher photocathode voltage required ($>6$ kV).
\item [5)] Shorter relative lifetime of GaAsP.
\item [6)] Temperature dependent avalanche gain in the APD.
\end{itemize}
For 4), one needs to take care of possible sparks (discharge) between neighbouring channels inside the camera.
For 5), although bombardment tests in the lab proved that the expected lifetime is $>10$ years, it is important to check this predictions in the field test (see \cite{RefHPD}).
For 6), in addition to the cooling system of the camera an additional temperature compensation circuit was implemented (see \cite{RefHPD}). The resulting temperature stability needs to be tested in real camera during the operation.

\subsection{The cluster module}
\begin{figure}[t]
\centering
  \centering
  \includegraphics[width=2in]{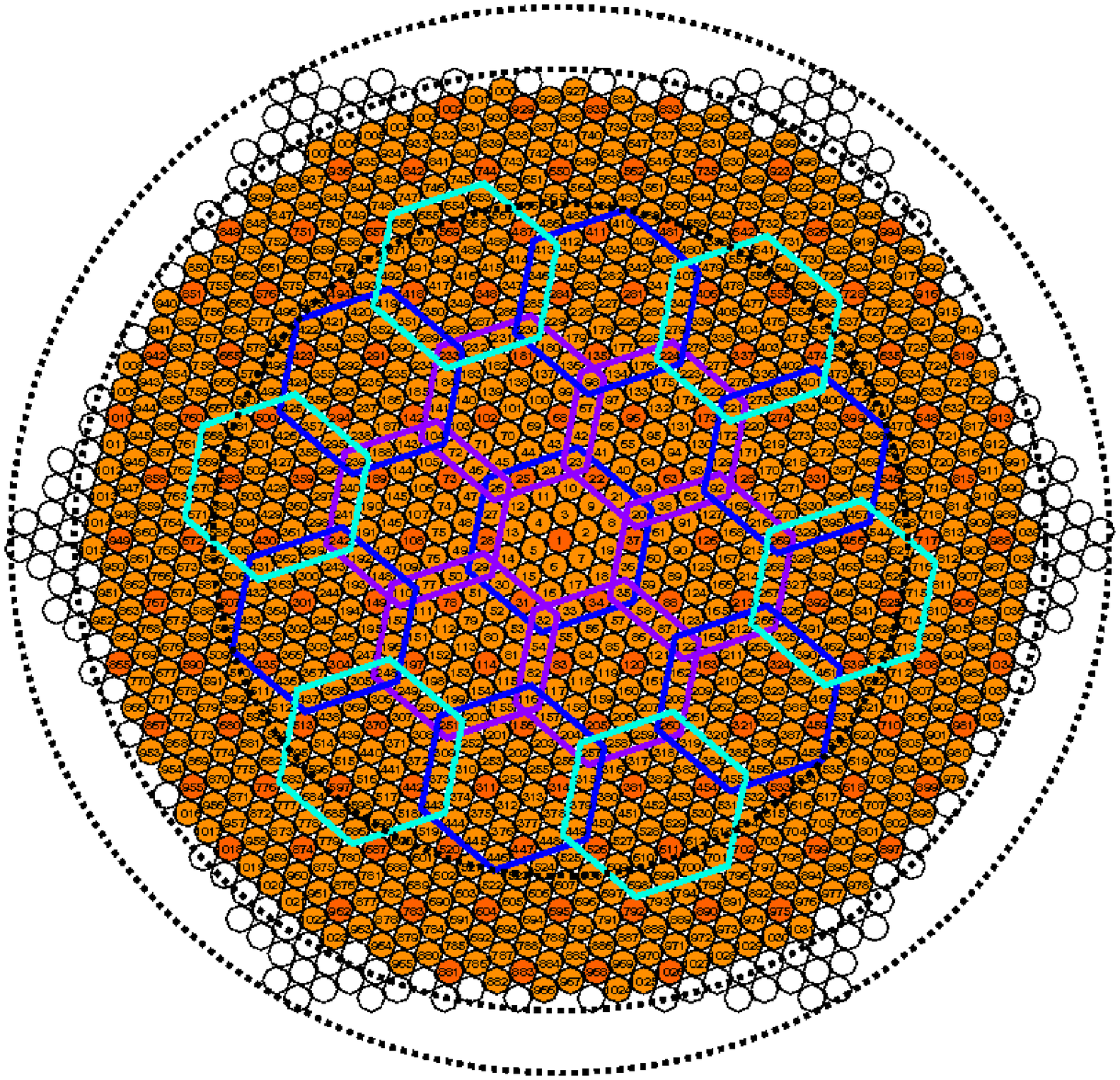}
  \includegraphics[width=2in]{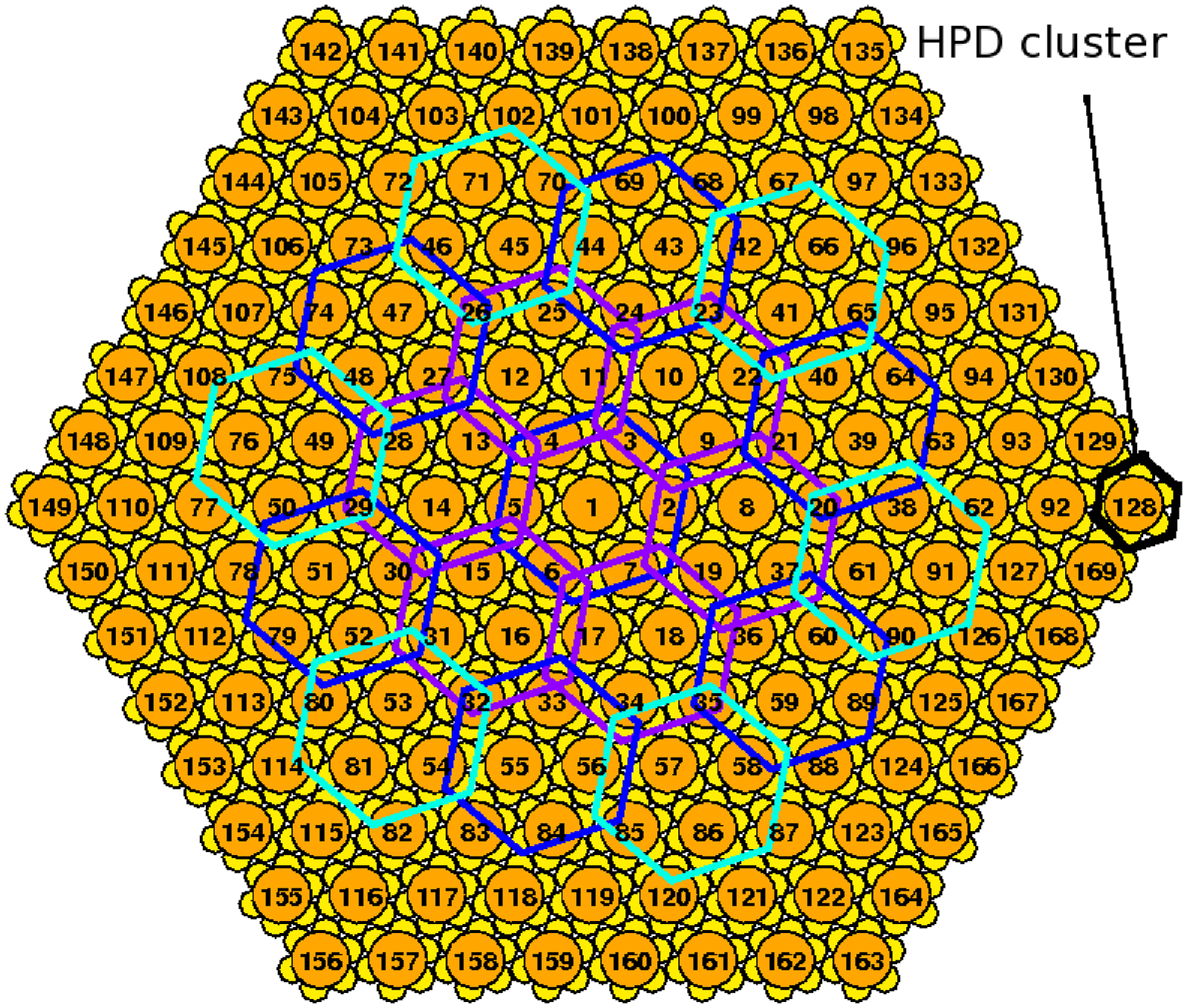} 
  \caption{Structure of the MAGIC-II camera. It consists of 163 PMT clusters. One cluster consists of 7 pixels. The HPD cluster is implemented at the right corner of the camera, which is not used by any PMT cluster.
  }
  \label{FigCam}
 \end{figure}

\begin{figure}[!t]
\centering
   \includegraphics[width=3.in]{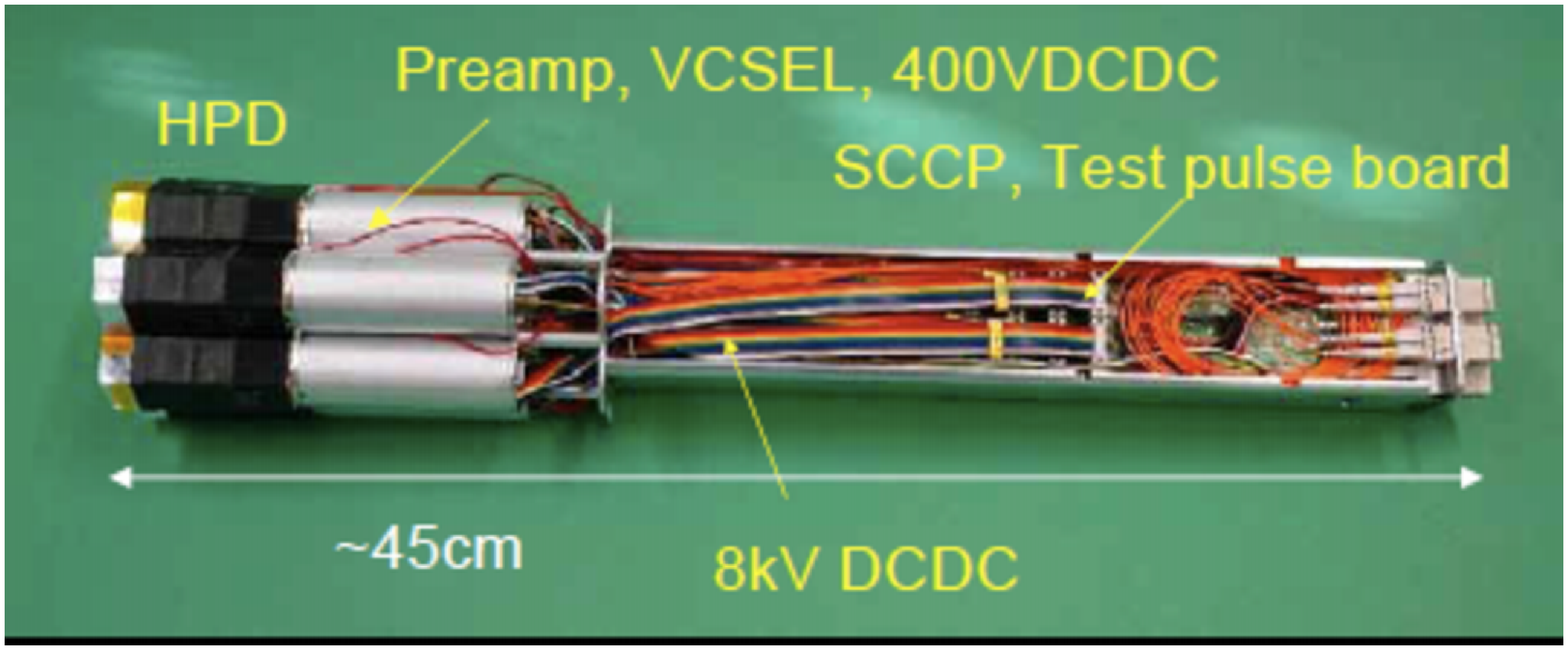}\label{fig2c}
   \includegraphics[width=1.5in]{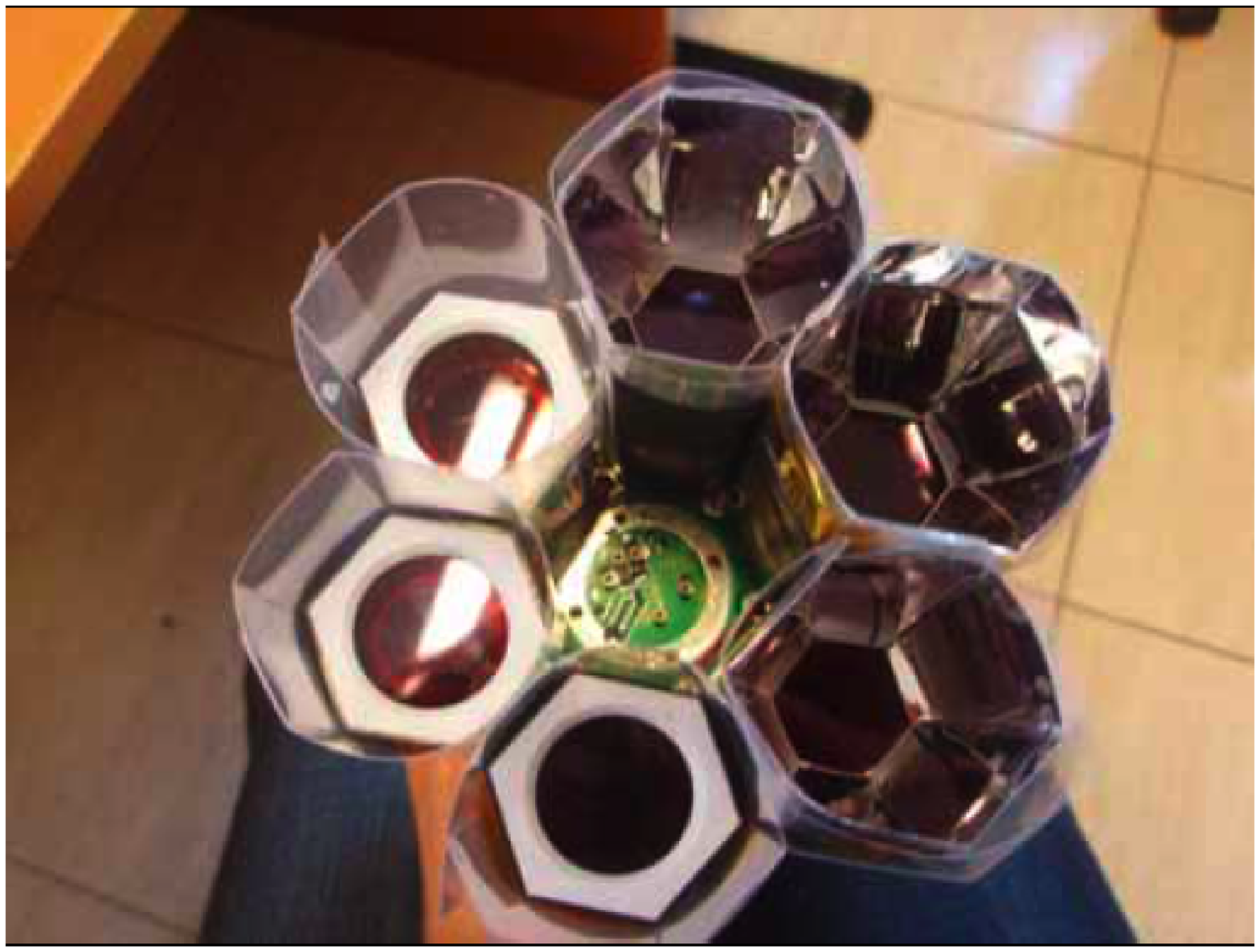} \label{fig3}
   \includegraphics[width=1.5in]{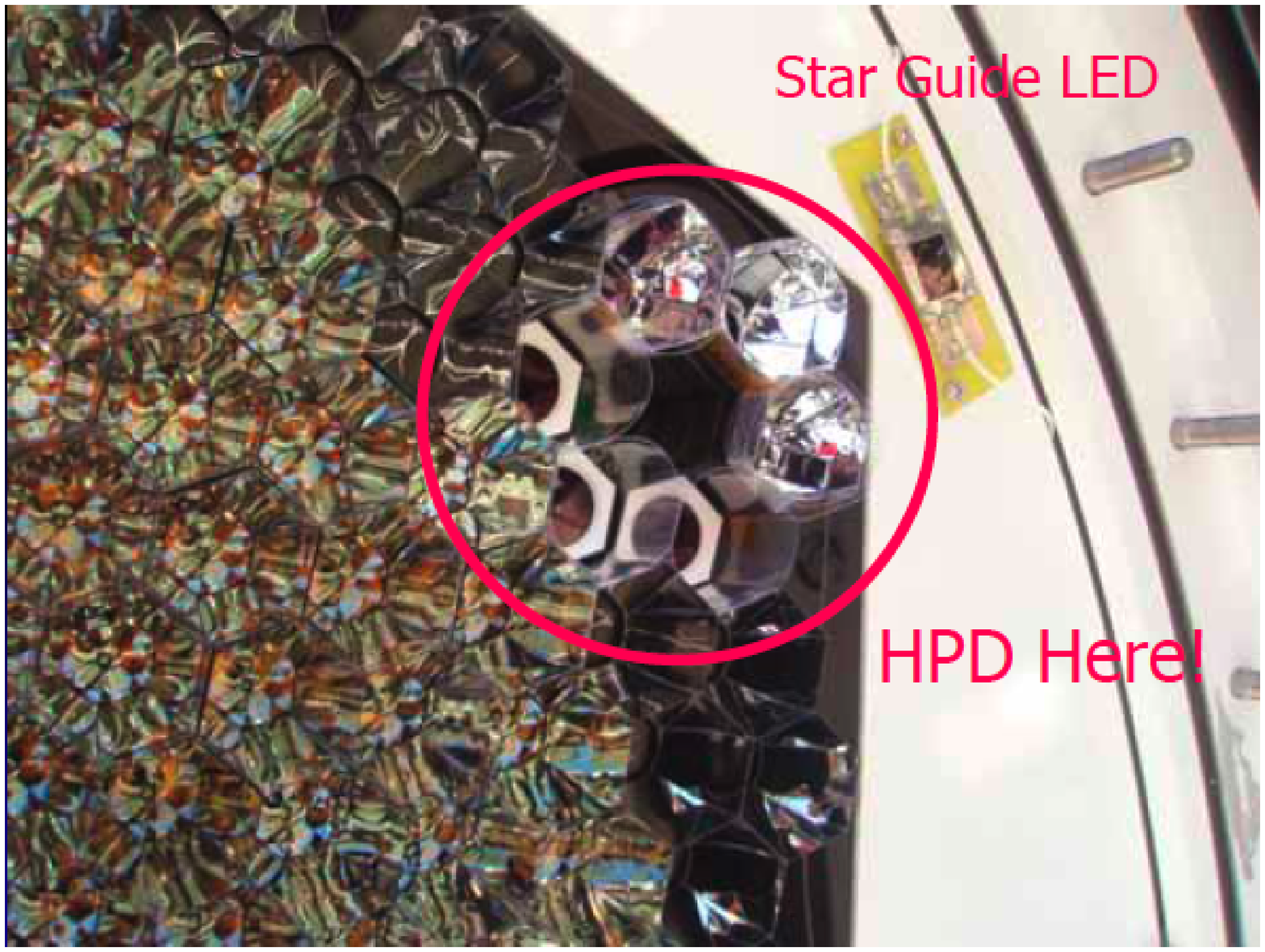} \label{fig3}
  
   \caption{Photographs of the HPD cluster. The central pixel is missing. Three pixels have
the dielectric Winston cone surrounded by a Capton foil, while the other three have only the Capton foil.}
   \label{FigHPDcluster}
 \end{figure}

The layout of the MAGIC-II camera is structured as follows:
7 PMTs make one hexagonal cluster. The cluster unit includes the high voltage power supplies, preamplifiers, readout circuits etc. for these PMTs. The camera body can hold up to 169 clusters, while only 163 PMT clusters (some of which have less than 7 PMTs) are installed to form a circular field of view (see Fig. \ref{FigCam}). This design allows to substitute individual cluster units with different detector modules. One prototype cluster with HPDs was build as shown in Fig. \ref{FigHPDcluster}. The cluster was primary installed in the middle right corner of the camera body. For mechanical reasons the central pixel of the cluster was not assembled in the prototype, i.e., the HPD cluster consists of 6 HPDs. The cluster contains one adjustable power supply with up to $-8$ kV, commonly used for the photocathodes of all HPDs. The APD bias voltages are supplied by individual supplies with voltages up to 500 V. Despite the power supplies readout circuits and temperature sensors are installed in the cluster.

\subsection{Winston cones made of dielectric foil}
In order to enlarge the field of view of each HPD and reduce the dead area between the pixels a Winston cone is attached to the entrance window of the photocathode. The Winstone cone design is adapted from the PMT clusters. For the PMTs aluminized Mylar foil is used. The same foil could not be used for the HPDs because it is electrically conductive. The HPD photocathode voltage is $\sim 6-7$ times higher than in a PMT and electrical discharges (sparks) between the photocathode and the Winstone cone, being at different potential, would appear. In the lab spark events with this foil could be observed using HPDs, while with PMTs they are rarely seen. 
Therefore we had to use dielectric material for the Winston cone's of the HPD cluster. Due to mechanical reasons only half of the HPDs in the prototype are assembled with Winstone cones (see Figure \ref{FigHPDcluster}). In order to eliminate further possible sparks between the HPDs additional Capton foil is surrounding each unit. It has to be mentioned that the dielectric material used for the Winston cone has very
poor reflectivity in UV range ($\sim 20$\% at 350 nm). This reduces drastically the collection efficiency of Cherenkov photons in this prototype. However, it shows the suppression of sparks. Materials with reflectivity higher than $\sim 95$\% at 300 nm are available and will be implemented in the near future.

\section{Results}
\subsection{Pulse shape}
\begin{figure}[h]
\centering
  \includegraphics[width=1.8in]{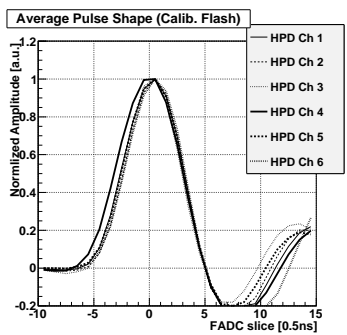}
  \caption{Pulse shape of the 6 HPDs recorded during the field test.
    The light source is a calibration laser with an width of $\sim 1$ ns.
    The FWHM of the output pulses are 2.7-3.0 ns.
    }
  \label{FigPulse}
 \end{figure}

The pulse shapes recorded during the field test are shown in Fig. \ref{FigPulse}. The FWHMs are ranging from 2.7 to 3.0~ns, which is slightly better than in PMTs.
The undershoot and overshoot are caused by the MAGIC readout system. This effects will be eliminated after the currently ongoing upgrade of the readout system.

\subsection{Shower images and sparks}
\begin{figure}[h]
\centering
   \includegraphics[width=3.in]{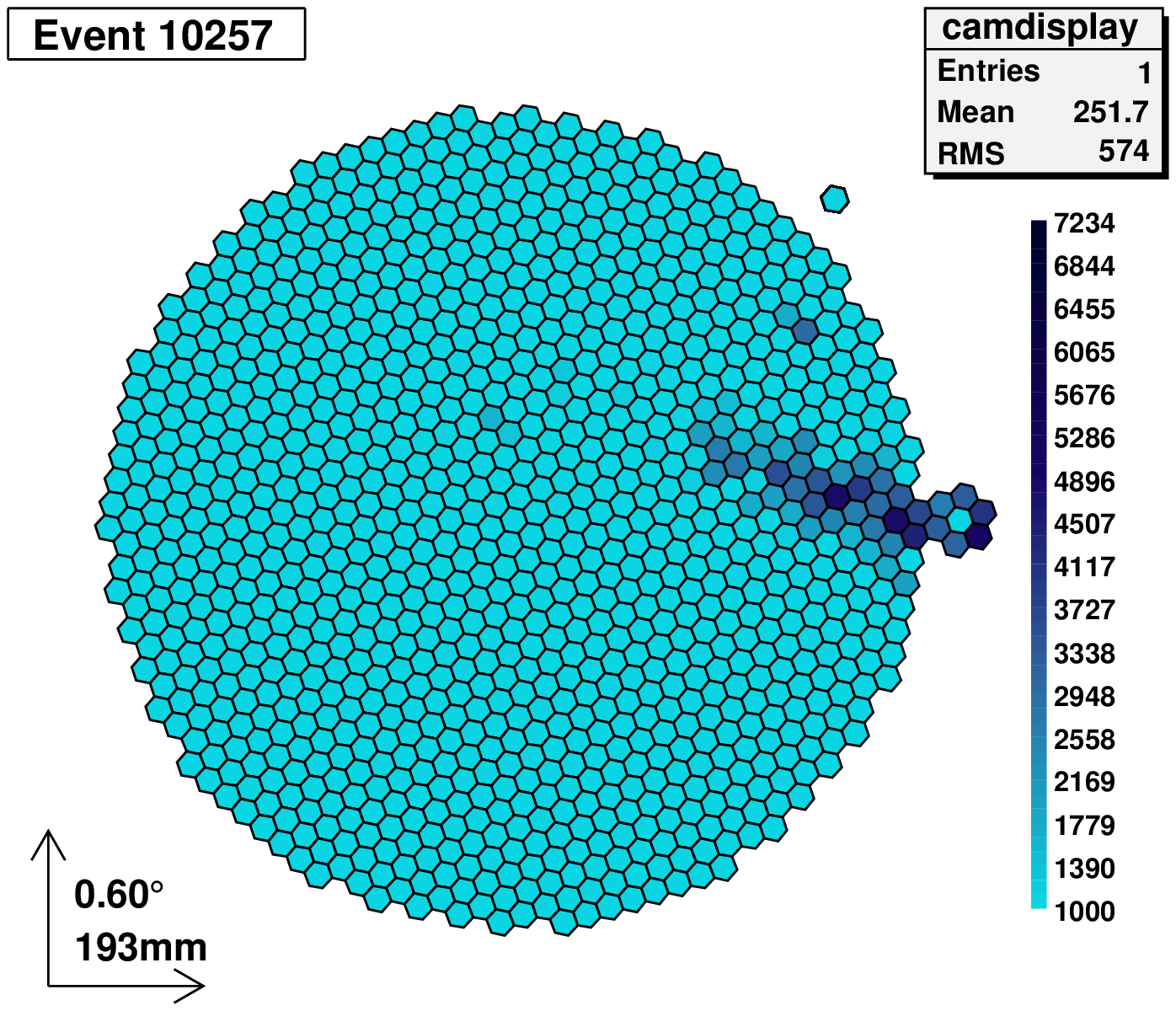}\label{fig2a}
   \includegraphics[width=3.in]{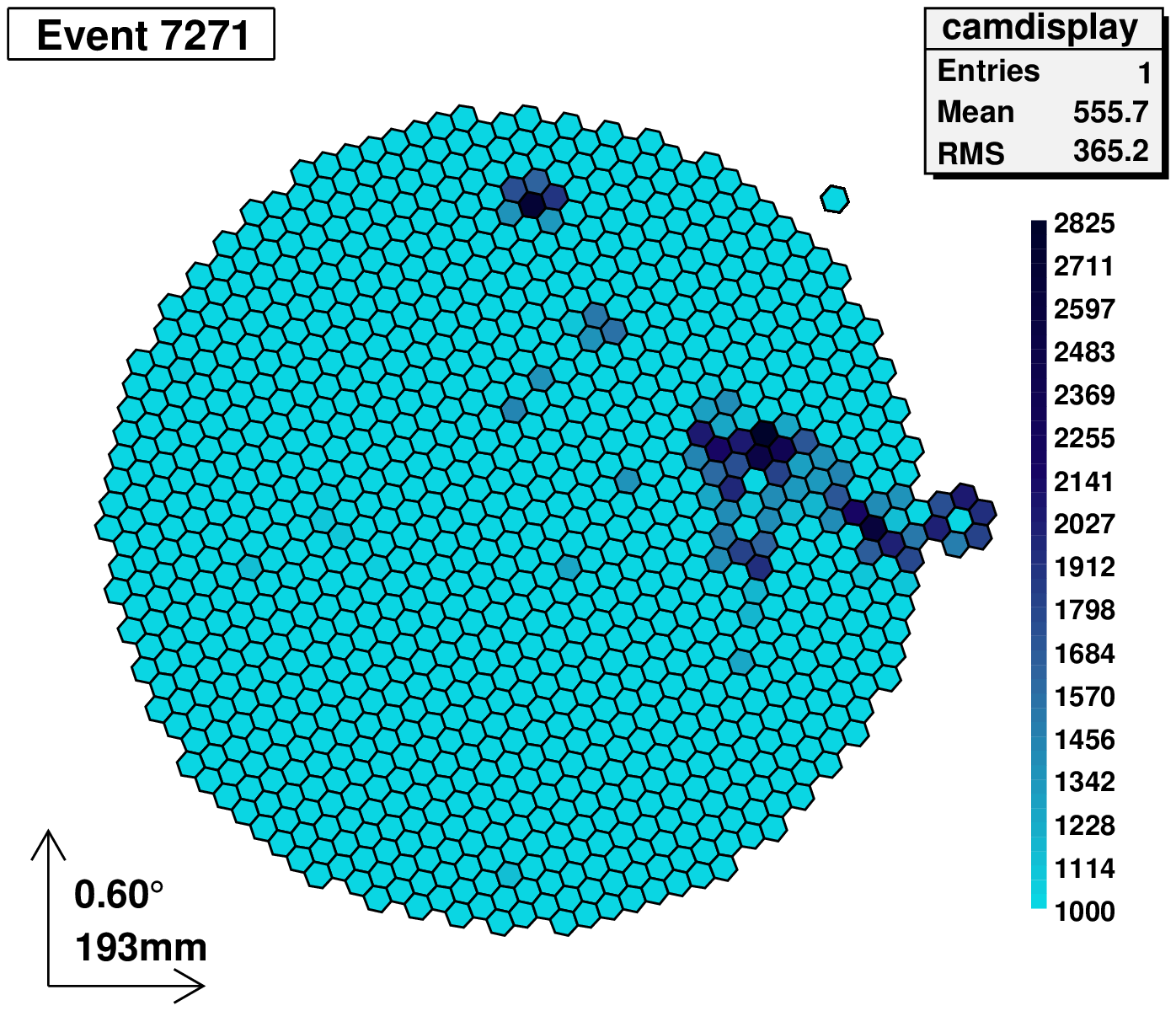} \label{fig3a}
   \includegraphics[width=3.in]{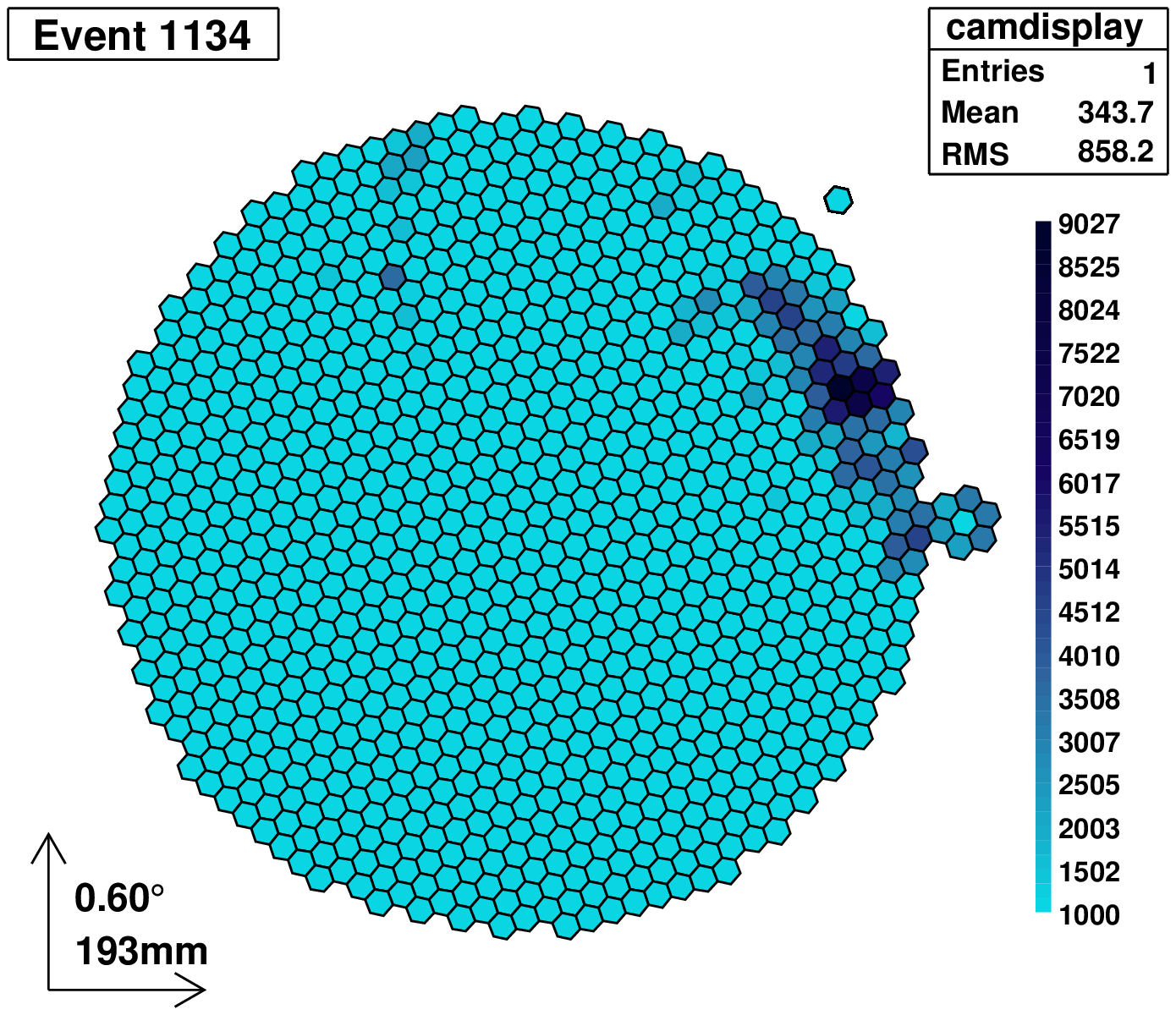} \label{fig3b}
   \caption{Some examples of air shower images in which the HPDs participate.
            }
   \label{FigImages}
 \end{figure}

During the field test the HPDs are switched on only during dark time (PMTs are operated even 
under moderate Moon light and twilight conditions). Some examples of shower images
are shown in Figure \ref{FigImages}. With the photocathode voltage of $-6.5$~kV no spark events were found in the data. 
 
\subsection{Degradation of the photocathode}

\begin{figure}[h]
\centering
  \includegraphics[width=2in]{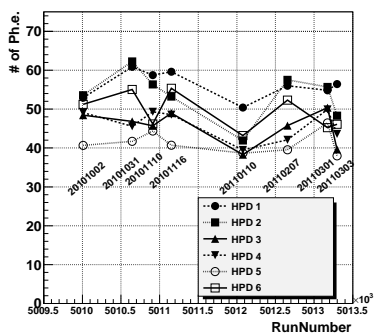}
   \caption{Number of detected ph.e's in the calibration events as a function of time. For the last half year no significant degradation is seen. The F-factor method does not have a precision better than 10\% in the estimated 
number of ph.e's.
            }
\label{FigNphe}
\end{figure}

In order to estimate the degradation of the photocathode regular calibration events are used.
Several times during the night dedicated calibration runs are taken. The calibration box is installed at the telescope dish in front of the telescope camera and uses a laser with a wavelength of 350 nm. The laser pulses illuminate the telescope camera uniformly at a frequency of 200 Hz.
In addition interleaved calibration events at a rate of 25~Hz are recorded during the data taking.
Taking into account the low reflectivity or absence of the Winston cones the pulse intensity in the HPDs correspond to typically a level of 50 ph.e..
One can estimate the mean number of ph.e ($N_{phe}$) for the calibration events using the so-called F-factor method \cite{RefFfactor}, i.e., with the following formula:
\begin{eqnarray}
N_{phe} = F^2 \mu^2/(\sigma_{sig}^2 - \sigma_{ped}^2)
\end{eqnarray}
where $F$, $\mu$, $\sigma_{sig}$, and $\sigma_{ped}$ are
the F-factor (1.05 for the HPDs), the mean signal charge, 
the RMS of the signal charge distribution, and the RMS of the pedestal charge distribution.
The results are shown in Fig. \ref{FigNphe}. 
Over half year scale no significant degradation is seen. This method has a precision of  $\sim 10$\% and more precise measurements, especially the quantum efficiency response, need to be done by dismounting the cluster and remeasuring it in the lab.

\subsection{Stability of the temperature}

\begin{figure}[h]
\centering
  \includegraphics[width=2in]{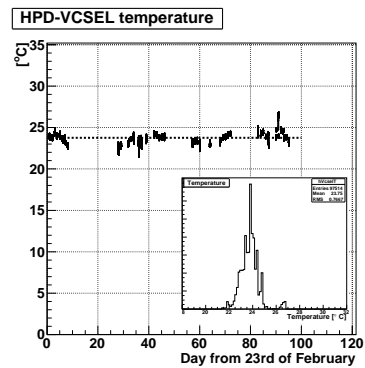}
\caption{Temperature near the VCSEL in the HPD cluster. It is very stable around 
24 degree. The variance of the avalanche gain can be neglected.}
\label{FigTemp}
\end{figure}

The avalanche gain has a temperature dependence of $-2$\%/$^\circ$C.
The implemented compensation circuit suppress it to $0.3$\%/$^\circ$C \cite{RefHPD},
but it is still important to stabilize the temperature within $\pm 3.3^\circ$C
so that the variance of the gain is less than 1\%. Fig. \ref{FigTemp} shows
that this condition was successfully fulfilled thanks to the efficient cooling system
of the MAGIC camera.

\section{Conclusions and perspectives}

The field test was very successful. The recorded signals fulfill the expectations and no sparks were found. Degradation of the quantum efficiency was not seen using the calibration events. However, the precision of this method is not good enough and dedicated measurements need to be done in the lab.
The temperature of the HPD cluster is well regulated. The variation of the avalanche gain is negligible.

This study was focussing mainly on possible disadvantages using HPDs. It results that the disadvantages can be controlled and are not problematic. The next step is to study the advantages using HPDs in IACT cameras. First  non-conductive Winston cones with high reflectivity will be used. A dielectric foil with 95\% reflectivity from 300 nm to 850 nm has already been produced.
Thanks to the high photon detection efficiency of HPDs the energy resolution and the background rejection power, especially for the low energy regime, can largely be improved. The very low after-pulsing rate of HPDs will help to lower the trigger energy threshold. These studies can be done using a group of HPD clusters in the MAGIC camera.

\subsection*{Acknowledgement}
We would like to thank the Instituto de Astrof\'{\i}sica de
Canarias for the excellent working conditions at the
Observatorio del Roque de los Muchachos in La Palma.
The support of the German BMBF and MPG, the Italian INFN,
the Swiss National Fund SNF, and the Spanish MICINN is
gratefully acknowledged. This work was also supported by
the Marie Curie program, by the CPAN CSD2007-00042 and MultiDark
CSD2009-00064 projects of the Spanish Consolider-Ingenio 2010
programme, by grant DO02-353 of the Bulgarian NSF, by grant 127740 of
the Academy of Finland, by the YIP of the Helmholtz Gemeinschaft,
by the DFG Cluster of Excellence ``Origin and Structure of the
Universe'', by the DFG Collaborative Research Centers SFB823/C4 and SFB876/C3,
and by the Polish MNiSzW grant 745/N-HESS-MAGIC/2010/0.


\clearpage

\end{document}